\documentclass[11pt,nofootinbib,preprint]{revtex4}
\usepackage{mathrsfs}
\usepackage{graphicx}
\usepackage{amsmath}
\usepackage{amsfonts}
\usepackage{amssymb}
\usepackage{color}

\usepackage{epsfig}
\usepackage{CJK}
\usepackage{graphicx}
\usepackage{epsfig}
\usepackage{eepic}
\usepackage{bbm}
\usepackage{dcolumn}
\usepackage{bm}
\usepackage{ulem}
\usepackage{tikz}

\newcommand{\omits}[1]{}

\def\bc{\begin{center}}

\def\ec{\end{center}}
\def\be{\begin{eqnarray}}
\def\ee{\end{eqnarray}}

\definecolor{dyellow}{rgb}{1.,0.8,.0}
\definecolor{myblue}{rgb}{.1,.1,.7}
\definecolor{dcyan}{rgb}{.0,.6,.6}
\definecolor{cyan}{rgb}{0.4,1.0,1.0}
\definecolor{dmagenta}{rgb}{0.6,0.0,0.6}
\definecolor{brown}{rgb}{0.6,0.2,0.}
\definecolor{darkblue}{rgb}{.0,.0,0.5}
\definecolor{darkred}{rgb}{0.75,0.0,0.0}
\definecolor{orange}{rgb}{1.,.6,.0}
\definecolor{dorange}{rgb}{0.8,.4,.0}
\definecolor{green}{rgb}{0.0,1.0,0.0}
\definecolor{darkgreen}{rgb}{0.0,0.6,0.0}
\definecolor{purple}{rgb}{.4,.0,.4}
\definecolor{lightgrey}{rgb}{0.7, 0.7, 0.7}
\definecolor{grey}{rgb}{0.4, 0.4, 0.4}

\def\tphi{\tilde \phi}

\newcommand{\nc}{\newcommand}
\nc{\rnc}{\renewcommand} \nc{\ket}[1]{\left | \, #1 \right \rangle}
\nc{\bra}[1]{\left \langle #1 \, \right |}
\nc{\ua}{\uparrow} \nc{\da}{\downarrow}

\nc{\braket}[2]{\langle\, #1\,|\,#2\,\rangle}
\nc{\half}{\frac{1}{2}}

\nc{\prj}{\mathcal{P}} \nc{\hilb}{\mathcal{H}}
\nc{\pth}{\mathcal{C}} \nc{\inprod}[2]{\braket{#1}{#2}}
\nc{\upket}{\ket{\uparrow}} \nc{\downket}{\ket{\downarrow}}
\nc{\upbra}{\bra{\uparrow}} \nc{\downbra}{\bra{\downarrow}}

\begin{document}


\title{Surface growth approach for bulk reconstruction in the AdS/BCFT correspondence}

\author{Xi-Hao Fang$^1$} \email{fangxh9@mail2.sysu.edu.cn}
\author{Fang-Zhong Chen$^1$} \email{chenfzh7@mail2.sysu.edu.cn}
\author{Jia-Rui Sun$^{1}$} \email{sunjiarui@mail.sysu.edu.cn}

\affiliation{${}^1$School of Physics and Astronomy, Sun Yat-Sen University, Guangzhou 510275, China}



\begin{abstract}
In this paper, we extend the surface growth approach for bulk reconstruction into the AdS spacetime with a boundary in the AdS/BCFT correspondence. We show that the geometry in the entanglement wedge with a boundary can be constructed from the direct growth of bulk extremal surfaces layer by layer. In addition, we find that the surface growth configuration in BCFT can be connected with the defect multi scale entanglement renormalization ansatz (MERA) tensor network. Furthermore, we also investigate the entanglement of purification within the surface growth process, which not only reveals more refined structure of entanglement entropy in the entanglement wedge but also suggests a selection rule for surface growth in the bulk reconstruction.
\end{abstract}

\pacs{04.62.+v, 04.70.Dy, 12.20.-m}

\maketitle

\section{Introduction}
There are different approaches in finding the quantum theory of gravity, one of the most promising approaches is the AdS/CFT correspondence or gauge/gravity duality developed from string theory. The AdS/CFT correspondence reveals that a $n$-dimensional gravitational system in the bulk asymptotically AdS spacetime is equivalent to a $(n-1)$-dimensional CFT on the timelike AdS boundary~\cite{Maldacena:1997re,Gubser:1998bc,Witten:1998qj}, which provides new perspectives and powerful tools to study strongly coupled field theories as well as quantum gravity. One of the most important progresses made in the studying of the AdS/CFT correspondence is the holographic interpretation of the entanglement entropy, i.e., the entanglement entropy between subsystem $A$ and its complement $B$ in a boundary CFT can be denoted by the area $A_\gamma$ of the minimal or extremal surface $\gamma$ (which is homologous to the boundary $\partial A$) in the bulk AdS spacetime, called the Ryu-Takayanagi (RT) formula of the holographic entanglement entropy (HEE)~\cite{Ryu:2006bv,Ryu:2006ef,Hubeny:2007xt,Rangamani:2016dms}
\be\label{rt}S_{A}= \frac{A_\gamma}{4G_N},\ee
where $G_N$ is the Newtonian constant. The formula in Eq.(\ref{rt}) establishes the connection between entanglement entropy of boundary quantum fields and the geometry of bulk gravity, which has been shown to play a key role in the bulk reconstruction, namely, using the information of the operators in the boundary CFT to reconstruct the bulk AdS gravity~\cite{bao1904,roy1801,faulkner1704,Harlow:2016,cotler1704}. This is just the emergent phenomenon of gravity indicated by the AdS/CFT correspondence and the more general gauge/gravity duality~\cite{Maldacena:1997re,Witten:1998qj}. An important progress in studying the bulk reconstruction in holography was the introduction of new approach from quantum manybody systems known as multi-scale entanglement renormalization ansatz (MERA) of tensor networks ~\cite{Vidal:2008,Swingle:2009bg,vidal:1812,vidal:2007}. MERA was originally served as a method to reduce computational complexity in solving the Schr\"{o}dinger equation of the lattice quantum manybody systems. It has been shown that the orientation of entanglement renormalization within MERA can be interpreted as the radial direction of the bulk AdS spacetime, which provides novel insights into the emergence of gravity and realization of AdS/CFT correspondence. Many different tensor networks have been studied to realize the models of holographic duality, see for example~\cite{Bao:2018pvs,Hung:2019zsk,Hayden:2016cfa,Vasseur:2018gfy}.
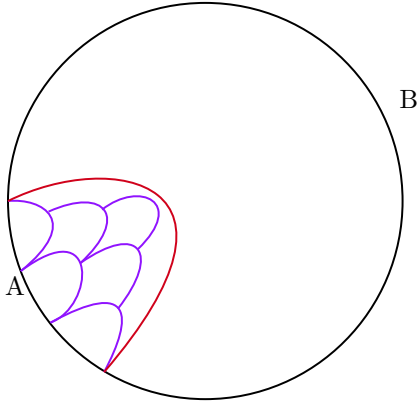
\begin{figure}[htbp]
	\begin{center}
		\tikzset{every picture/.style={line width=0.75pt}} 
		\begin{tikzpicture}[x=0.75pt,y=0.75pt,yscale=-1,xscale=1]
			
			\draw   (190,159) .. controls (190,103.77) and (234.55,59) .. (289.5,59) .. controls (344.45,59) and (389,103.77) .. (389,159) .. controls (389,214.23) and (344.45,259) .. (289.5,259) .. controls (234.55,259) and (190,214.23) .. (190,159) -- cycle ;
			\draw [color={rgb, 255:red, 144; green, 19; blue, 254 }  ,draw opacity=1 ]   (190,159) .. controls (207.45,157.35) and (227.45,169.35) .. (196.45,194.35) ;
			\draw [color={rgb, 255:red, 144; green, 19; blue, 254 }  ,draw opacity=1 ]   (196.45,194.35) .. controls (236.45,164.35) and (233.45,212.35) .. (211.45,220.35) ;
			\draw [color={rgb, 255:red, 144; green, 19; blue, 254 }  ,draw opacity=1 ]   (211.45,220.35) .. controls (253.03,197.7) and (253.45,217.35) .. (238.45,245.35) ;
			\draw [color={rgb, 255:red, 144; green, 19; blue, 254 }  ,draw opacity=1 ]   (210.45,164.35) .. controls (239.45,151.35) and (251.45,167.35) .. (226.45,190.35) ;
			\draw [color={rgb, 255:red, 144; green, 19; blue, 254 }  ,draw opacity=1 ]   (226.45,190.35) .. controls (256.45,169.35) and (267.45,185.35) .. (245.45,213.35) ;
			\draw [color={rgb, 255:red, 144; green, 19; blue, 254 }  ,draw opacity=1 ]   (237.45,163.35) .. controls (256.45,148.35) and (280.03,159.7) .. (255.45,183.35) ;
			\draw [color={rgb, 255:red, 208; green, 2; blue, 27 }  ,draw opacity=1 ]   (190,159) .. controls (216.52,145.85) and (249.52,143.27) .. (265.52,156.07) .. controls (281.51,168.86) and (279.99,197.03) .. (238.45,245.35) ;
			
			\draw (187,196) node [anchor=north west][inner sep=0.75pt]   [align=left] {A};
			\draw (386,101.6) node [anchor=north west][inner sep=0.75pt]   [align=left] {B};
		\end{tikzpicture}
	\end{center}	
	\caption{The surface growth scheme in the AdS spacetime, bulk extremal surfaces growing from subintervals in the subregion $A$ gives an efficient way to construct the bulk geometry and detect the fine structure of entanglement entropy in the entanglement wedge.}
\end{figure}

Recently, more and more evidences indicate that quantum entanglement is probably deeply connected with the essence of spacetime geometry and gravity, such as the subregion-subregion duality in the entanglement wedge reconstruction and the bulk reconstruction from boundary quantum error code~\cite{faulkner1806,Almheiri:2014lwa,Pastawski:2015qua,Penington:2019npb}. An important question is how to find general and more efficient approaches to reconstruct bulk geometry and gravitational dynamics from the information of boundary operators. Very recently, a new approach for bulk reconstruction named surface growth approach was proposed in \cite{Lin:2020thc,sun1912}. This scheme points out that the bulk AdS spacetime can be constructed by growing of bulk extremal surfaces layer by layer from the boundary, which is similar to the Huygens' principle of wave propagation. It has been shown that the surface growth scheme can correspond to a MERA-like tensor network called the one shot entanglement distillation (OSED) tensor network~\cite{Bao:2018pvs}, and has been explicitly realized by direct growth of bulk extremal surfaces in the asymptotically AdS$_3$ spacetime~\cite{Yu:2020zwk}.

On the other hand, when the boundary CFT contains additional spatial boundary, its holographic duality is a bulk asypotically AdS spacetime with a brane in it, which is called the AdS/BCFT correspondence~\cite{Takayanagi:2011zk,Cardy:2004hm,Fujita:2011fp,Chu:2017aab}.
The BCFT contains several new properties than the ordinary CFT, one of which is that the entanglement entropy will receive contributions from the boundary. From the HEE description, the bulk extremal surface can end both on the AdS boundary and the brane in bulk AdS spacetime~\cite{Fujita:2011fp,Takayanagi:2011zk}, which gives a natural picture of boundary to bulk propagation and can be viewed as process of bulk reconstruction from entanglement entropy. In addition, the AdS/BCFT correspondence has been used as an equivalent description for entanglement island in the braneworld viewpoint, see, for example~\cite{Almheiri:2019hni,Chen:2020uac,Chen:2020hmv,Lin:2022aqf,Miao:2023unv,Jeong:2023lkc,Bhattacharya:2021jrn,Bhattacharya:2021nqj,Geng:2020fxl,Geng:2020qvw,Geng:2023qwm}. Therefore, it is interesting to further study the entanglement entropy and bulk reconstruction in the asymptotically AdS spacetime with boundary. In the present paper, we extend the surface growth approach for bulk reconstruction into the AdS/BCFT correspondence. We show that the entanglement wedge with a boundary on the brane can be effectively constructed by the growth of bulk extremal surfaces and discuss the possible connection between the surface growth configuration and the defect MERA and defect OSED tensor networks. In addition, we also analyze the entanglement of purification by investigating the entanglement wedge cross section in the surface growth process, which gives the fine structure of entanglement entropy in the entanglement wedge and suggests a selection rule for surface growth in the bulk reconstruction in the AdS/BCFT correspondence.

This paper is organized as follows: In Sec.~\ref{hee bcft}, we briefly review of the HEE in the AdS/BCFT correspondence. In Sec.~\ref{sg bcft} and Sec.~\ref{dfc MERA}, we study the surface growth scheme in AdS$_5$/BCFT$_4$ correspondence both in the homogeneous and inhomogeneous subregion cases and discuss the connection between the surface growth configuration with the defect tensor network. In Sec.~\ref{sg eop}, we study entanglement of purification in AdS$_3$/BCFT$_2$  by analyzing the entanglement wedge cross section in the surface growth process. Finally we give the conclusion and discussion.

\section{HEE in the AdS/BCFT correspondence}\label{hee bcft}
The AdS$_{d+1}$ spacetime in Poincar\'{e} coordinates is given by
\begin{equation}
ds^2=\frac{L^2}{z^2}\left(-dt^2+dz^2+dx_idx^i\right),
\end{equation}
where $i=1,\ldots,d-1$ and $L$ is the curvature radius. The asymptotically timelike boundary of the AdS spacetime is located at $z=\epsilon \to 0$. Considering the entanglement entropy of a strip-like subregion $A$ with $x_1\equiv x\in[-l/2,l/2]$, the bulk codimension-2 static minimal surface is $x=x(z)$, its explicit form is determined by the Euler-Lagrange equation for its area functional $A=L_0^{d-2}\int \frac{L^{d-1}}{z^{d-1}}\sqrt{1+{x'}^2}dz$, which is
\begin{equation}\label{ELeq}
x'=\frac{1}{\sqrt{\frac{z_*^{2d-2}}{z^{2d-2}}-1}},
\end{equation}
where we have assumed that $x_a$ for $a=2,\ldots,d-2$ are in the box region with length to be $L_0$. The turning point of the bulk extremal surface is $z_*$, which satisfies $x'(z_*)= +\infty$ and
\begin{equation}
\int_0^{z_*}  x'(z)dz=\int_0^{z_*}\frac{dz}{\sqrt{\frac{z_*^{2d-2}}{z^{2d-2}}-1}}=\frac{l}{2},
\end{equation}
When choosing the background spacetime to be pure AdS$_5$ spacetime without boundary,
\begin{equation}
z_*=\frac{l\Gamma(\frac{1}{6})}{2\sqrt{\pi}\Gamma(\frac{2}{3})}.
\end{equation}
The equation of the minimal surface $\gamma$ is a hypergeometric function:
\begin{equation}
x=\frac{z_*}{4}\left(\frac{z}{z_*}\right)^4 F\left(\frac{1}{2},\frac{2}{3},\frac{3}{5},\left(\frac{z}{z_*}\right)^6\right).
\end{equation}

When considering a bulk spacetime which is a portion of the AdS spacetime, there is a boundary $Q$ which can be regarded as a brane in the bulk spacetime. While from the perspective of boundary CFT side, there is a spatial boundary $P$ which is homologous to $Q$, which is called the BCFT. The existence of a boundary has nontrivial effects both on the gravity and CFT sides. In the case of a bulk minimal surface encountering a $d$ dimensional brane $Q$, which can be regarded as the end-of-world (EOW) brane in the bulk AdS spacetime, and in the Poincar\'{e} coordinate, the location of a flat boundary (brane) can be determined as by $x=z\tan \theta$, where $\theta+\pi/2$ is the angle between the brane and the AdS boundary. For a strip-like bulk minimal surface $\gamma$ intersecting with the boundary on point $P_0$ with coordinate $(z_0\tan\theta,z_0)$ in the $(x,z)$ plane, the boundary condition requires that $\gamma$ must be perpendicular to the boundary $Q$. In other words, the unit normal vectors of the boundary $\hat{n}_Q$ should be perpendicular to the unit normal vectors of the minimal surface  $\hat{n}_\gamma$ where  $\hat{n}_Q\cdot \hat{n}_\gamma=0$, which gives $x'(z)|_Q=-\cot\theta$. By integrating the equation of motion with the boundary condition,
\be
\int_\epsilon^{z_*}x'(z)dz+\int_{z_*}^{z_0}x'(z)dz=z_0\tan\theta+l,
\ee
the turning point $z_*$ is obtained 
\be
\frac{l}{z_*}=\frac{2\sqrt{\pi}\Gamma (\frac{2}{3})}{\Gamma (\frac{1}{6})}+\frac{B\left( \cos ^2\theta ;\frac{2}{3},\frac{1}{2} \right)}{6}\,\,-\tan \theta \left( \cos \theta \right) ^{\frac{1}{3}},
\ee
in which the subregion $A$ is chosen as $x_1\equiv x\in[-l,0]$, $B(z,a,b)$ represents the incompatible Beta function and $\epsilon$ is the UV cutoff of the BCFT. Subsequently, by further dividing the coordinate $x$ of the strip-like subregion $A$ into multiple intervals, one can construct the surface growth from these subintervals in subregion $A$ layer by layer. 

\begin{figure}[htbp]
	\begin{center}\label{fig2}
		\includegraphics[width=1\linewidth,height=9cm,clip]{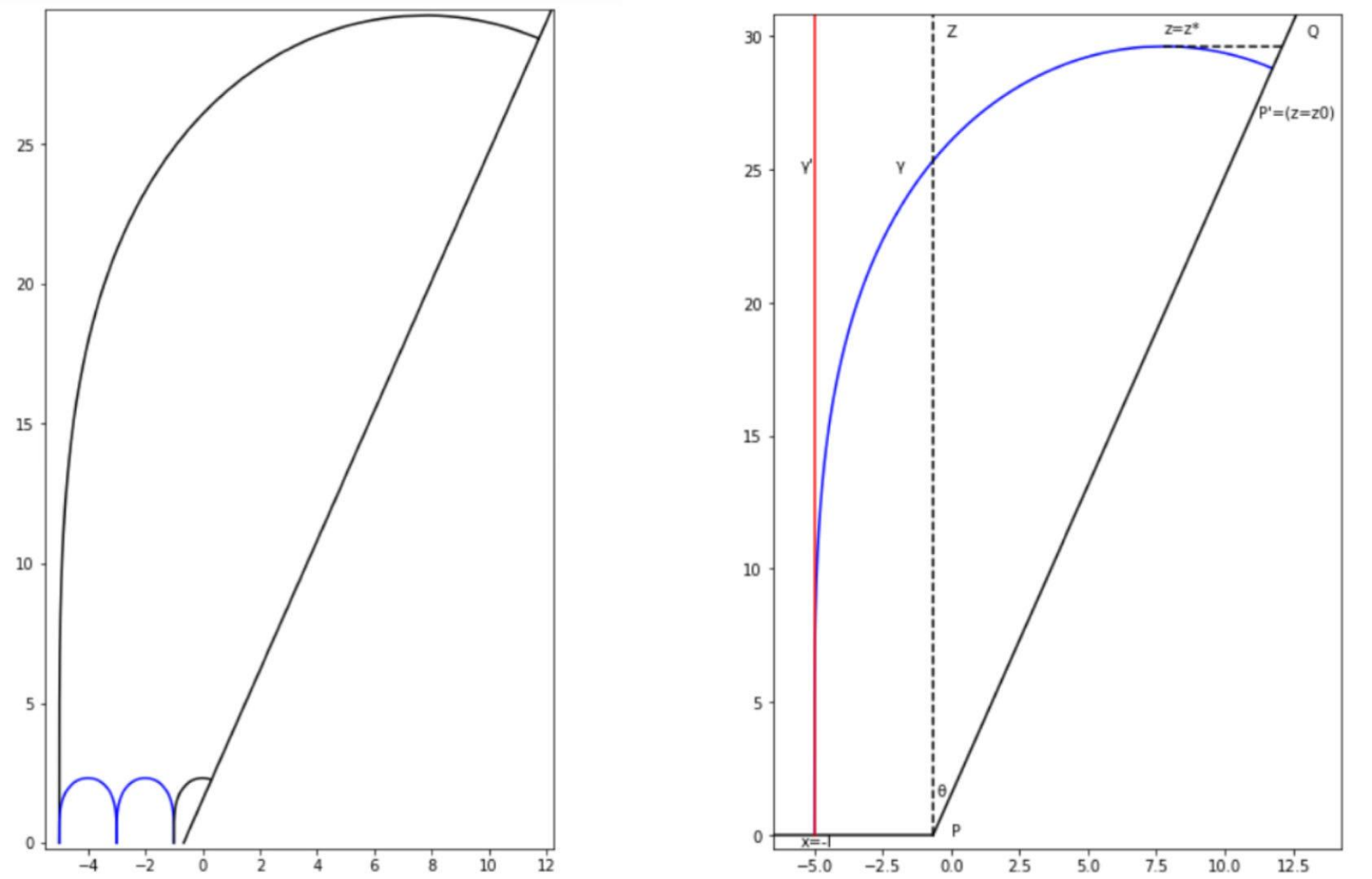}
		\caption{Left: the first layer of strip-like bulk minimal surfaces in subregion $A$ in the AdS$_5$ spacetime with boundary, $\theta =\frac{\pi}{4}$, the size of the left two subintervals in $A$ is $ \Delta x_1=2$, and the rest one is $ \Delta x_1=0.18$, and $L=1$. The blue surface represent the type $ \mathcal{B} $ minimal surface, the black surface represent the boundary (defect) type $ \mathcal{A} $ minimal surface which will end on the boundary.  Right: the bulk minimal surface which intersects with the boundary with vertical boundary condition, the blue line denotes the type $ \mathcal{A} $ minimal surface, the red line represents the asymptotic curve parallel to the $ z $-axis, the subregion A is chosen as $x_1\in (-5,0)$. Both figures are drawn in the $x_1-z$ plane.}
		\label{fig2}
	\end{center}	
\end{figure}
\section{Surface growth scheme in AdS$_5$ spacetime with boundary}\label{sg bcft}
To generalize the surface growth scheme into asymptotically AdS spacetime with a boundary $Q$ and see how the presence of spatial boundary will affect the configuration of growing minimal surfaces. Firstly, let us consider a half-infinite strip boundary $Q$ in the bulk AdS spacetime, where the boundary of the BCFT is located at $x=x_P$. The strip-like subsystem $A$ is divided into several subintervals as $\{x_N,x_{N+1}\}$, $\left( N\in \mathbb{N} \right)$, where each endpoint of the subinterval is the starting point of the bulk minimal surface $\left. \mathrm{\gamma}_{N} \right|_{z=\epsilon}$. Due to the restriction imposed by the boundary $Q$ on the half strip, the bulk minimal surface corresponding the interval which contains $x_P$ will end on (perpendicular to) the brane $Q$. Therefore, naturally, the bulk minimal surfaces are classified into two types: type $ \mathcal{A} $ surfaces which intersect on the boundary $Q$ and have a vertical intersection, while type $ \mathcal{B} $ surfaces have both endpoints attaching on the boundary CFT, see Fig.~\ref{fig2}.

After obtaining the first layer of bulk minimal surfaces in the strip-like subregion $A$ as shown in Fig.~\ref{fig2}, one can further construct the geometry in a larger region in the entanglement wedge through the growth of bulk minimal surfaces. Following the approach in~\cite{Lin:2020thc}, in the same spirit of the Huygens' principle of wave propagation, one layer of bulk minimal surface on any cutoff surface in AdS spacetime can be viewed as the starting point for bulk minimal surfaces of the next layer. This provides the initial conditions for the Euler-Lagrange equation (\ref{ELeq}) satisfied by the bulk minimal surfaces. For example, considering two neighboring minimal surfaces $\gamma_N$ and $\gamma_{N+1}$ on two type $ \mathcal{B} $ adjoining intervals $\left\{ x_N,x_{N+1} \right\} ,\left\{ x_{N+1},x_{N+2} \right\} $, respectively, one can choose appropriate starting/ending points on each of them to determine the next layer of bulk minimal surfaces. 

\begin{figure}[htbp]
	\begin{center}\label{fig4}
   \includegraphics[width=0.5\linewidth,height=8.5cm,clip]{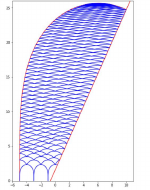}
   \caption{The surface growth scheme in AdS$_5$ spacetime with a boundary, $\theta =\frac{\pi}{4}$ and $L=1$. The subregion $A$ is divided into three subintervals, the left two have $ \Delta x_1=2$ and the rest one has $ \Delta x_1=0.18$. There are three kinds of starting/ending points for the growth of bulk extremal surfaces, one is on the bulk extremal surface of the subregion $A$, namely, the red RT surface, one is the turning point $z_*$, and the rest is a new one on the boundary $Q$. The number of growing layers are 50.}
   \label{fig4}
  \end{center}	
  \end{figure}

To fully construct the geometry in the entanglement wedge of subregion $A$ with boundary from the growth of bulk minimal surfaces, apart from starting/ending points on the turning point $z_*$ and the points on the bulk extremal surface of subregion $A$, additional starting/ending points on the boundary $Q$ is necessary. This selection of bulk extremal surfaces not only guarantees the continuity between the preceding and succeeding layers but also ensures the seamless propagation of information from the boundary CFT to the interior of bulk AdS spacetime. More explicitly, when dealing with a type $ \mathcal{A} $ surface that intersects with the boundary, its influence extends to the next layer as long as it continues to intersect with the boundary. This leads to a situation where the initial condition of $x'(z)$ is determined by both the turning point of the previous layer, $z_{*}$, and a point on the boundary $Q$, denoted as $\left( x_0, z_0 \right) $. These initial conditions are adopted to define the RT surface in the subsequent layer. Besides, apart from bulk minimal surfaces in the first layer, subsequent layers are no longer directly connected to the entanglement entropy of subregions in the boundary CFT. Instead, they will reflect the way how the information of the boundary CFT will propagate into the deep AdS spacetime. The bulk extremal surfaces are generalized RT surfaces denoted as $\bar{\gamma}$ and they provide a more refined description of entanglement entropy in subregion-subregion duality in the entanglement wedge reconstruction~\cite{Lin:2020thc}. The surface growth scheme involves iterating this procedure layer by layer, eventually the entire entanglement wedge of the subregion $A$ will be filled, which gives a direct and efficient way to construct the bulk geometry in the entanglement wedge, as shown in Fig.~\ref{fig4}.

\begin{figure}[htbp]
	\centering
	\begin{minipage}{6cm}\hfill
		\centering
		\includegraphics[width=1\linewidth,height=6.0cm,clip]{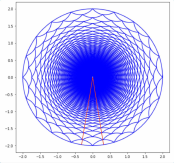}
	\end{minipage}
	\begin{minipage}{6cm}\hfill
		\centering
		\includegraphics[width=0.95\linewidth,height=6.0cm,clip]{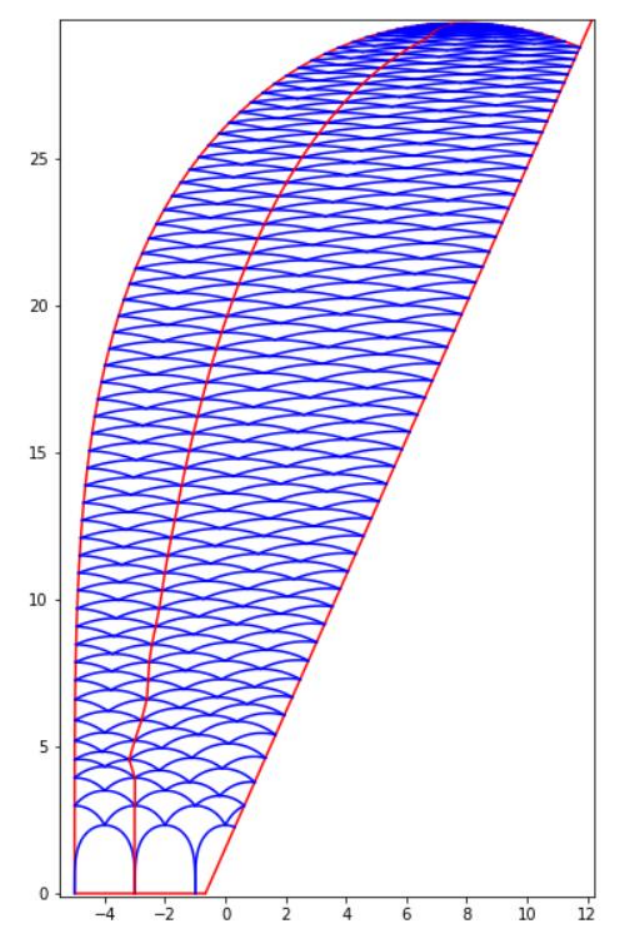}
	\end{minipage}	
\caption{Surface growth of strip-like bulk extremal surfaces in AdS spacetimes with and without the influence of boundary effects, distinct properties of growing areas are shown. Left: The region between the two red curves represents a bunch of growing surfaces and the boundary subregions are equally partitioned, the areas of each layer decrease monotonically. The bulk spacetime is AdS$_3$ with $L=1$. Right: The region between the left two red curves (surfaces) represents a bunch of growing surfaces, the areas of each layer do not exhibit a strictly decreasing trend when going into the bulk AdS spacetime with boundary. The bulk spacetime is AdS$_5$ with $L=1$.}
\label{fig42}
\end{figure}

Note that in the presence of a boundary $Q$, the area of bulk extremal surface within each layer of surface growth does not exhibit a strictly decreasing trend solely with respect to the radial coordinate $z$, which is distinct with the monotonicity property in surface growth process without boundary, as depicted in both the left and right figures of Fig. \ref{fig42}. The appearance of this new phenomenon is a consequence of the boundary effect, namely, the existence of additional spatial boundary $Q$ imposes constraints on the growth rules, leading to a situation where surfaces within the same layer do not possess identical horizontal positions. Consequently, this effect results in non-monotonic variations in surface area across layers. However, for regions near the apex of the outer entanglement wedge, the areas of growing extremal surfaces still exhibit a monotonically decreasing property, which is similar to the surface growth process without boundary. This new property highlights the important role of boundary conditions in shaping the spatial distribution of growing extremal surfaces within layers in the surface growth scheme.

\begin{figure}[htbp]
	\centering
	\begin{minipage}{6cm}
	  \centering	  \includegraphics[width=1\linewidth,height=7cm,clip]{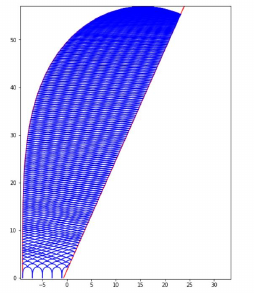}
	  \label{fig51}
	\end{minipage}
	\begin{minipage}{6cm}
	  \centering
	  \includegraphics[width=1\linewidth,height=7cm,clip]{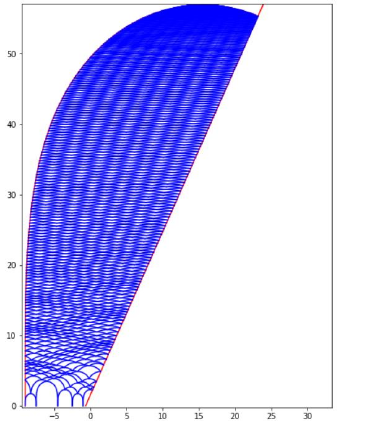}
	  \label{fig52}
	\end{minipage}

\caption{Comparison of surface growth of strip-like bulk extremal surfaces between homogenous (Left) and inhomogenous (Right) boundary subintervals in AdS$_5$/BCFT$_4$ with $\theta =\frac{\pi}{3}$ and $L=1$. The subregion $A$ is divided into five subintervals. Left: The surface growth of homogenous subintervals, the left four subintervals have $ \Delta x_1=2 $, the rest one has $ \Delta x_1=0.18 $, and number of growing layers is 500. Right: The surface growth for boundary subregions with inhomogenous subintervals, and the growing layers are also 500.}
	\label{fig5}
  \end{figure}

Furthermore, we can also consider the surface growth for the case in which the boundary subregions have inhomogenous size. As illustrated in Fig.~\ref{fig5}, similar to the case of surface growth without boundary, although the growing surfaces are inhomogeneous initially, the configuration becomes close to the homogenous case with a sufficient number of growth layers.

\section{Relation between the surface growth scheme and the defect MERA}\label{dfc MERA}
In the previous section, we successfully extended the surface growth scheme into the AdS spacetime with a boundary, namely, the translational symmetry is broken at the boundary. It was shown that a set of strip-like bulk extremal surfaces can fill the entanglement wedge of a given boundary subregion layer by layer, and two distinct types of extremal surfaces will emerge during the surface growth process. A comparative analysis between surface growth scheme with and without a boundary is also made. On the other hand, recall that in the absence of boundary, the surface growth scheme can correspond to the MERA-like tensor network~\cite{Lin:2020thc}, thus it is natural to ask whether the surface growth scheme in AdS/BCFT duality can still correspond to some tensor network. To answer this question, note that from the geometric picture of the MERA, the surface growth scheme in AdS/BCFT duality is expected to correspond a MERA with boundary, which can be called the defect MERA~\cite{Evenbly:2009}.

To generalize the MERA of tensor network with boundary and especially its holographic duality, one method is called the minimal updates~\cite{Czech:2016nxc,Evenbly:2013tta}. The main idea of the minimal updates is that the initial system can naturally evolve into the defect system by modifying a segment of its inner structure located within the causal cone associated with defect or boundary in the lattice. Letting $H$ represents the Hamiltonian of a homogeneous many-body quantum state, and $H^{\#}$ denotes the system with any defect, then
\be
H^{\#}=H+H_d,
\ee
where $H_d$ represents the interaction term associated with a defect or boundary. With this Hamiltonian, the quantum state description of MERA undergoes corresponding changes to accommodate to the defect MERA, denoted as wave functions $|\varphi \rangle^{\#}$ and $|\varphi \rangle$ respectively. Then the causal cone of the defect can be defined through a consistent coarse-graining process.

\begin{figure}[htbp]

	\begin{minipage}{6cm}
 \includegraphics[height=7cm,clip]{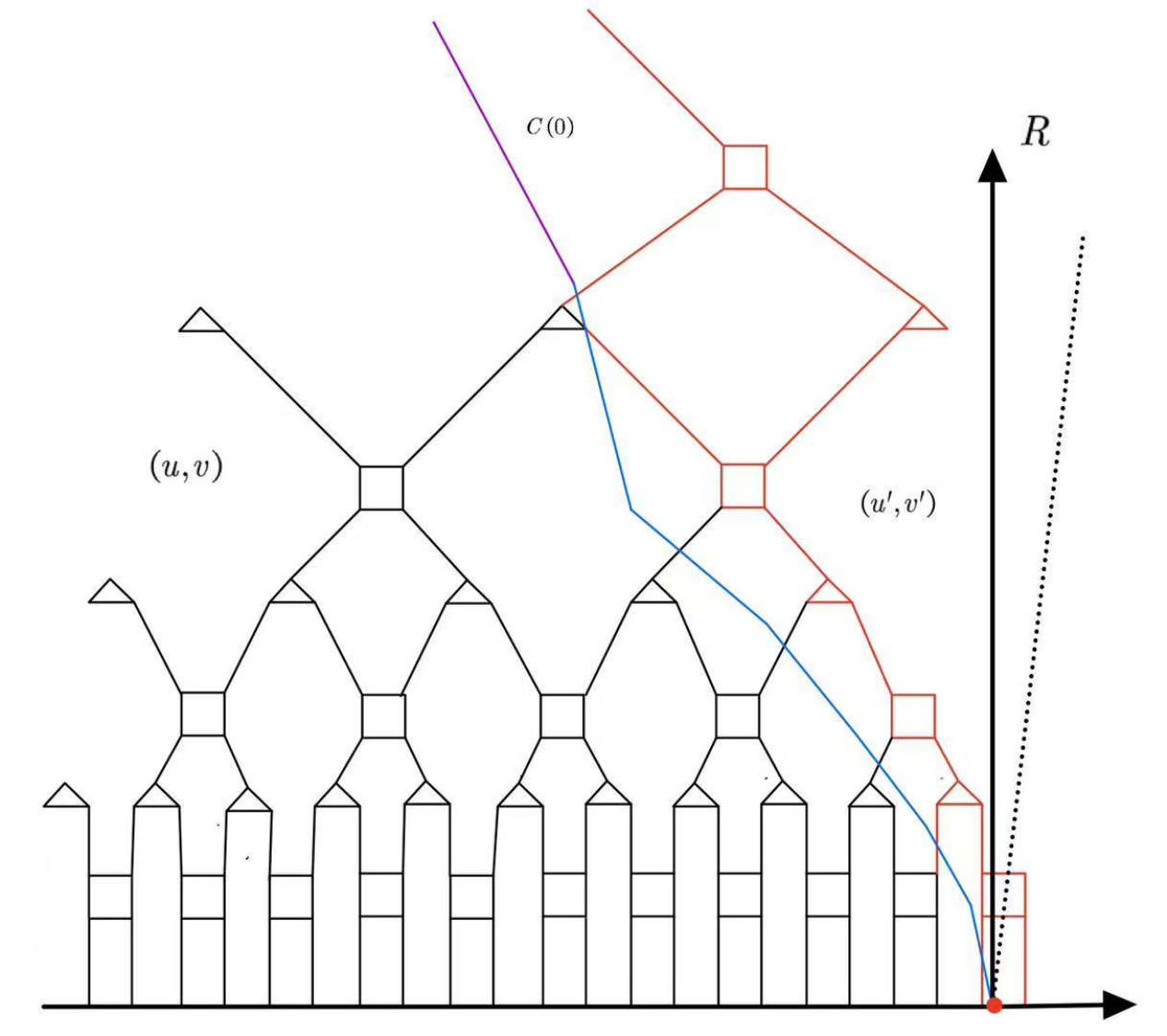}
		\label{43}
	\end{minipage}
\hspace{6.5em}
	\centering
	\begin{minipage}{6cm}
		\includegraphics[height=7.5 cm,clip]{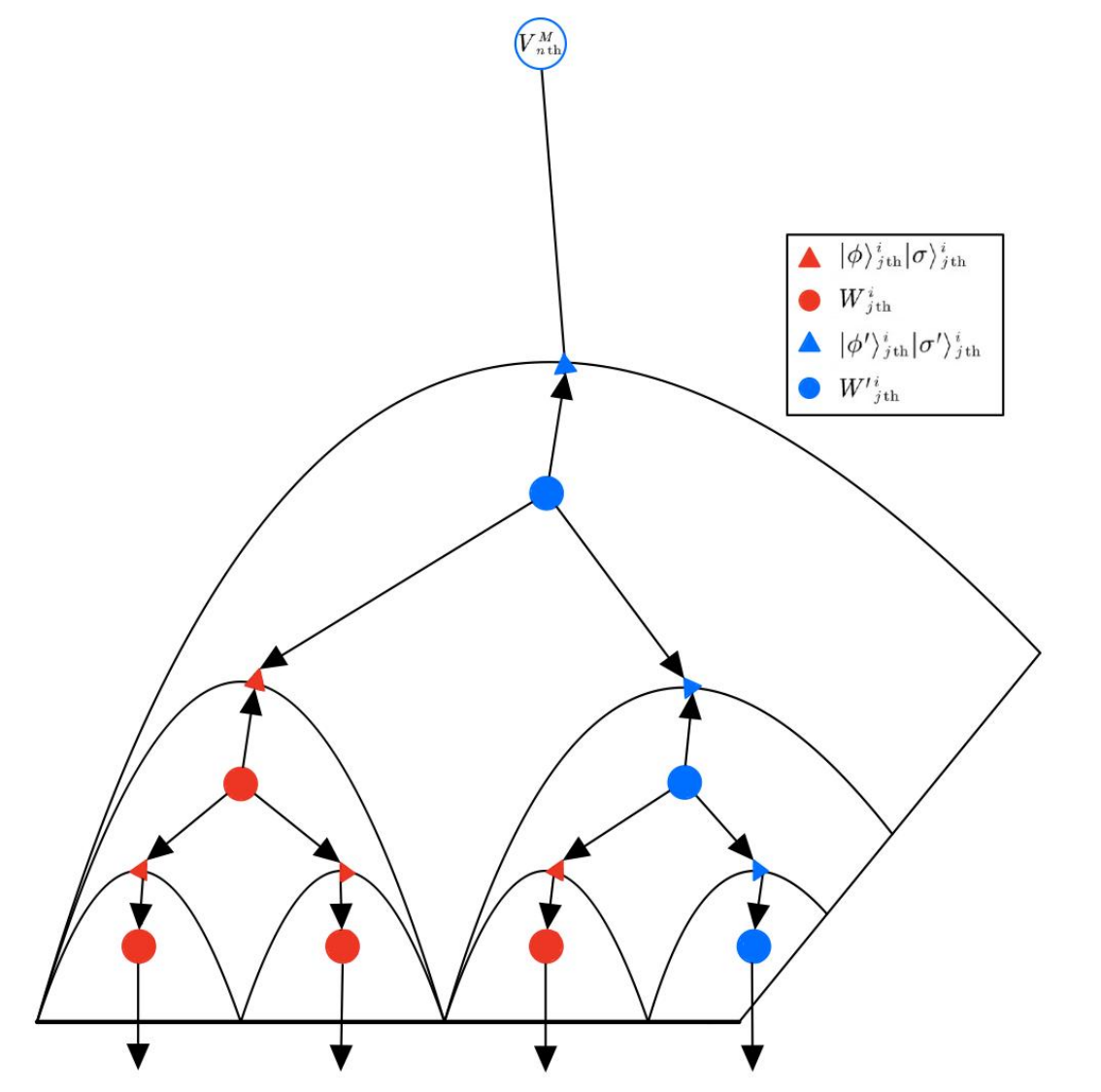}
		\label{44}
	\end{minipage}
\centering
\caption{The structure of defect MERA and product OSED process. Left: blue line denotes the boundary of the casual cone $C(0)$ associated with the defect or boundary in the lattice, $R$ is the direction of renormalization. The red square and triangle denote the update isometry and disentangler tensor inside and outside the causal cone respectively. Right: due to boundary effect, the modified OSED tensor network contains two representation of approximate quantum state production, corresponding to the boundary surface growth scheme.}
\label{fig6}
\end{figure}

Considering a one dimensional critical quantum system, in which the homogeneous Hamiltonian is $H=\sum_{r=-\infty}^{\infty}{h(r,r+1)}$, with constant coupling constant $h$. The MERA of this system contains a pair of tensors  $\left\{ u,v \right\} $, where $ u $ is the disentangeler  and $ v $ is coarse-grainer. According to the minimal updates, the defect MERA of the quantum state with scale invariance manifests two pairs of component tensors, denoted as $\left\{ u,v \right\} $ and $\left\{ u',v' \right\} $, describing regions outside and inside the causal cone $C(0)$ of the defect or boundary $D$, respectively. Then by combining the surface growth approach in~\cite{Lin:2020thc} and the method of minimal updates~\cite{Czech:2016nxc,Evenbly:2013tta}, it is reasonable to identify the surface growth configuration $\gamma$ and $\bar{\gamma}$ with tensor pairs $\left\{ u,v \right\} $ and $\left\{ u',v' \right\} $, as shown in Fig.~\ref{fig6}, in which $\left\{ u',v' \right\}  \Leftrightarrow \bar{\gamma}$ and $\left\{ u,v \right\}\Leftrightarrow \gamma$.

More explicitly, the state of the BCFT can be represented as a product of holographic tensor network state, i.e., the distilled maximal entanglement state
\begin{equation}
|\varPsi \rangle =W_{1\bar{\beta}\bar{\alpha}}^{A}W_{2\bar{\beta}\bar{\alpha}}^{\bar{A}}\phi ^{\alpha \bar{\alpha}}\sigma ^{\beta \bar{\beta}}=\left( W_1\otimes W_2 \right) \left( |\phi \rangle \otimes |\sigma \rangle \right) \otimes \left( W'_1\otimes W'_2 \right) \left( |\phi'\rangle \otimes |\sigma'\rangle \right),
\end{equation}
where $W_1$ and $W_2$ represent the isometry tensors which map the states $|\phi \rangle$ and $|\sigma \rangle$ into the reduced density matrix of boundary subregion $A$ and its complement $\bar{A}$, respectively. $|\phi \rangle$ is the dominant state and $|\sigma \rangle$ describes the quantum fluctuation. While $W'$, $|\sigma' \rangle$ and $|\sigma' \rangle$ are their counterparts within the causal cone $C\left( 0 \right) $. By repeating the above entanglement distillation procedure, the OSED defect tensor network corresponding to the surface growth with boundary can be constructed as in Fig.~\ref{fig6}, which gives a specific form of bulk reconstruction from surface growth in the AdS/BCFT correspondence, tensor network realization in such correspondence can also be found in \cite{Mori:2023swn}. In \cite{Mori:2022xec}, the authors presented a method similar to surface growth scheme in which the boundary is pushed toward the bulk.

\section{Entanglement of purification in surface growth scheme}\label{sg eop}
Since the surface growth process can divide the entanglement wedge into arbitrary subregions, which is expected to capture the fine structure the entanglement entropy in these subregions such as the entanglement contour~\cite{Vidal:2014aal,Mo:2023kym}. In addition, it was shown in~\cite{Lin:2020yzf} that the holographic duality of the entanglement purification~\cite{Nguyen:2017yqw,Takayanagi:2017knl} can also be derived from a specific surface growth scheme. Therefore, it is interesting to further investigate the entanglement of purification in surface growth scheme in the AdS/BCFT correspondence. The holographic duality of entanglement of purification can be described by the entanglement wedge cross section $\Sigma _{AB}$ as~\cite{Nguyen:2017yqw,Takayanagi:2017knl}
\be
E_P\left( \rho _{AB} \right) =\min \left[ \frac{\mathrm{Area}\left( \Sigma _{AB} \right)}{4G_N} \right],
\ee
where $A$ and $B$ are two boundary disjoint subsystems, the entanglement wedge $W$ are bounded by extremal surfaces of $A$ and $B$, and the connected extremal surfaces $\Sigma _A$ and $\Sigma _B$ of region $A\cup B$.

\begin{figure}
	\centering
	\centering
	\includegraphics[width=0.6\linewidth,height=6 cm,clip]{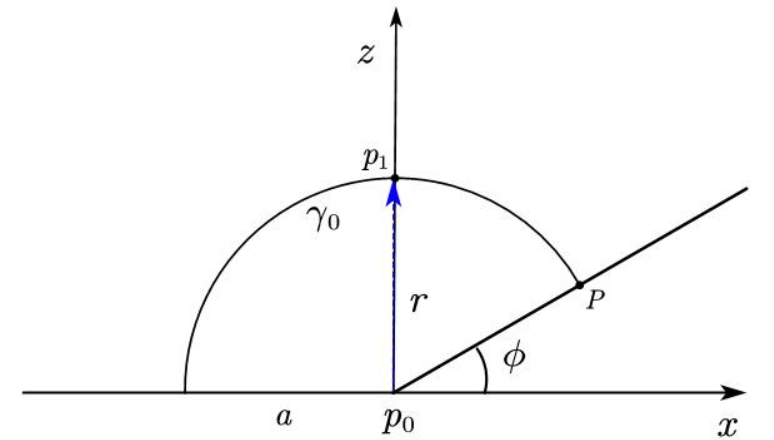}
	\caption{The entanglement wedge cross section of subregion $A$ (with length $a$) and boundary region $p_{0}P$ of a BCFT in AdS$_3$ spacetime. The blue arrow denotes the position of the entanglement wedge cross section $p_{0}p_{1}$ which connects the boundary point $p$ and the bulk RT surface, and $\phi =\frac{\pi}{6}$. The holographic description of the entanglement of purification corresponds to $p_0p_1$ when it is a minimal surface.}
	\label{fig61}
\end{figure}

Considering a BCFT case in AdS$_3$ spacetime as illustrated in Fig.~\ref{fig61}, the subregion $A$ and a boundary region $p_0P$ share a entanglement wedge, the entanglement wedge cross section $p_0p_1$ is defined as the surface between the endpoints of the boundary $p_0$ and the RT surface $\gamma_0$. To describe the entanglement of purification, $p_0p_1$ should be a minimal surface (namely, a geodesic). Note that the formula of geodesic distance $D$ between two spacetime points $(t_1,x_1,z_1)$ and $(t_2,x_2,z_2)$ in AdS$_3$ spacetime in Poincar\'{e} coordinate is
\be \label{eop}
\cosh\frac{D}{L}=\bigg|1+\frac{-(t_1-t_2)^2+(x_1-x_2)^2+(z_1-z_2)^2}{2z_1z_2}\bigg|,
\ee
and the static geodesic (namely, $t_1=t_2$) satisfies the equation of a circle, i.e., $(x-x_c)^2+(z-z_c)^2=r^2$ (where $(x_c,z_c)$ is the location of the center and $r$ is the radius of the circle, respectively). In Fig.~\ref{fig61}, setting the angle between $p_0p_1$ and $p_0P$ to be $\theta$, the radius of the geodesic circle is $r$, and the coordinate $p_0$ on the asymptotically AdS$_3$ boundary is $(0,\epsilon)$, then from eq.(\ref{eop}), the static geodesic distance between $p_0$ and $p_1$ satisfies
\be 
\cosh\frac{D_{p_0p_1}}{L}
=1+\frac{r^2\cos^2(\theta+\phi)+\left(r\sin(\theta+\phi)-\epsilon\right)^2}{2\epsilon r\sin(\theta+\phi)},
\ee
then $\frac{d}{d\theta}\cosh\frac{D_{p_0p_1}}{L}=0$ gives $\cos(\theta+\phi)=0$, i.e. $\theta+\phi=\frac{\pi}{2}$, namely, the entanglement of purification corresponds to the case in which $p_0p_1$ is perpendicular to $x$ axis.

Next, we will focus on studying the entanglement of purification in a specific surface growth configurations in the AdS$_3$/BCFT$_2$ correspondence via investigating the entanglement wedge cross section at each step of the surface growth process. The specific surface growth configuration is consisted of extremal surfaces which will end both on the previous one and the boundary, and the entanglement of purification are geodesics between the point $p_i$ and the next layer extremal surface $\gamma_i$. Note that in the process of surface growth, the starting and ending points of the next layers depend on various conditions such as the size and symmetry of the previous layers, as well as the boundary condition. Since the entanglement of purification is a minimal surface which contains minimal entanglement entropy from the BCFT side, we suggest that it will serve as a selection rule to chose the starting point of the growing minimal surfaces. 

\begin{figure}
	\centering
	\centering
	\includegraphics[width=0.6\linewidth,height=6 cm,clip]{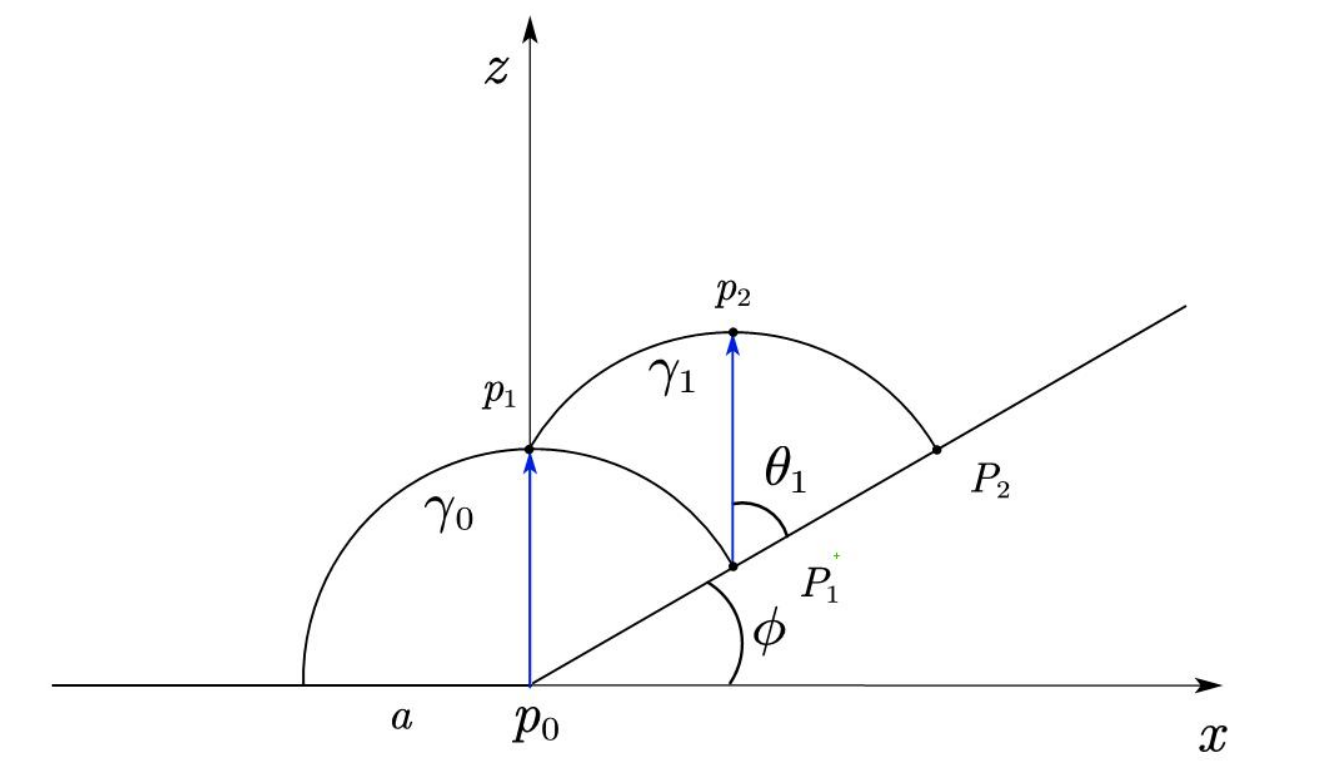}
	\caption{The entanglement wedge cross section in the second layer of growing minimal surface, the boundary angle is $\phi =\frac{\pi}{6}$, $ \gamma_{0} $ and $ \gamma_{1}$ denote the first and second layer extremal surfaces, respectively. The blue arrow denotes the geodesic between two surface, $ \theta_1 $ is the angle between the entanglement wedge cross section $P_1p_2$ and the boundary. }
	\label{fig62}
\end{figure}

For the two layers surface growth case, as shown in Fig.~\ref{fig62}, the closed surface $p_1P_1P_2p_2$ can be holographically interpreted as a pure state from the surface/state correspondence~\cite{miyaji:1503}, then choosing the geodesic curve $p_1P_1$ and the region $P_1P_2$ as the subsystems both in the bulk and on the boundary, respectively, the entanglement of purification in the second layer corresponds to minimize the entanglement wedge cross section $P_1p_2$, and the angle $\theta_1$ will be determined in the following multi-layers surface growth configuration.

In the surface growth process, from the second layer on, each layer of the extremal surface starts at the point $ p_{i} $ of the previous layer, meaning that each surface intersects both the boundary and the previous entanglement wedge cross section. Then the entanglement of purification in the next layer also follows the surface growth scheme and be calculated by finding the minimal surface between the point $ p_{i} $ and the next extremal surface $\gamma_{i+1}$. More explicitly, the ending point $P_i$ of the $i$-th extremal surface on the boundary naturally serves as the new boundary of the spacetime, we can choose location of $ P_i$ to be $\left( m_i,n_i \right)$, the angle between the entanglement wedge cross section $P_ip_{i+1}$ with the boundary is $ \theta_{i} $, and the radius of the geodesic circle $\gamma_i$ to be $r_i$. Then using eq.(\ref{eop}), the geodesic distance $D_{P_ip_{i+1} }$ between the boundary point $ P_{i} $ and the intersection point $p_{i+1}$ on the extremal surface $\gamma_i$ satisfies
\begin{equation}
	\cosh \frac{ D_{P_ip_{i+1} }}{L}=1+\frac{\left( r_i\cos (\theta_i+\phi)\right) ^2+\left( r_i\sin(\theta_i+\phi) \right) ^2}{2n_i\left(n_i+ r_i\sin(\theta_i+\phi)\right)}.\label{5.1}
\end{equation}
Then $\frac{d}{d\theta_i}D_{P_ip_{i+1} }=0$ also gives $\theta_i+\phi=\frac\pi 2$.

\begin{figure}
	\centering
	\centering
	\includegraphics[width=0.9\linewidth,height=7.5 cm,clip]{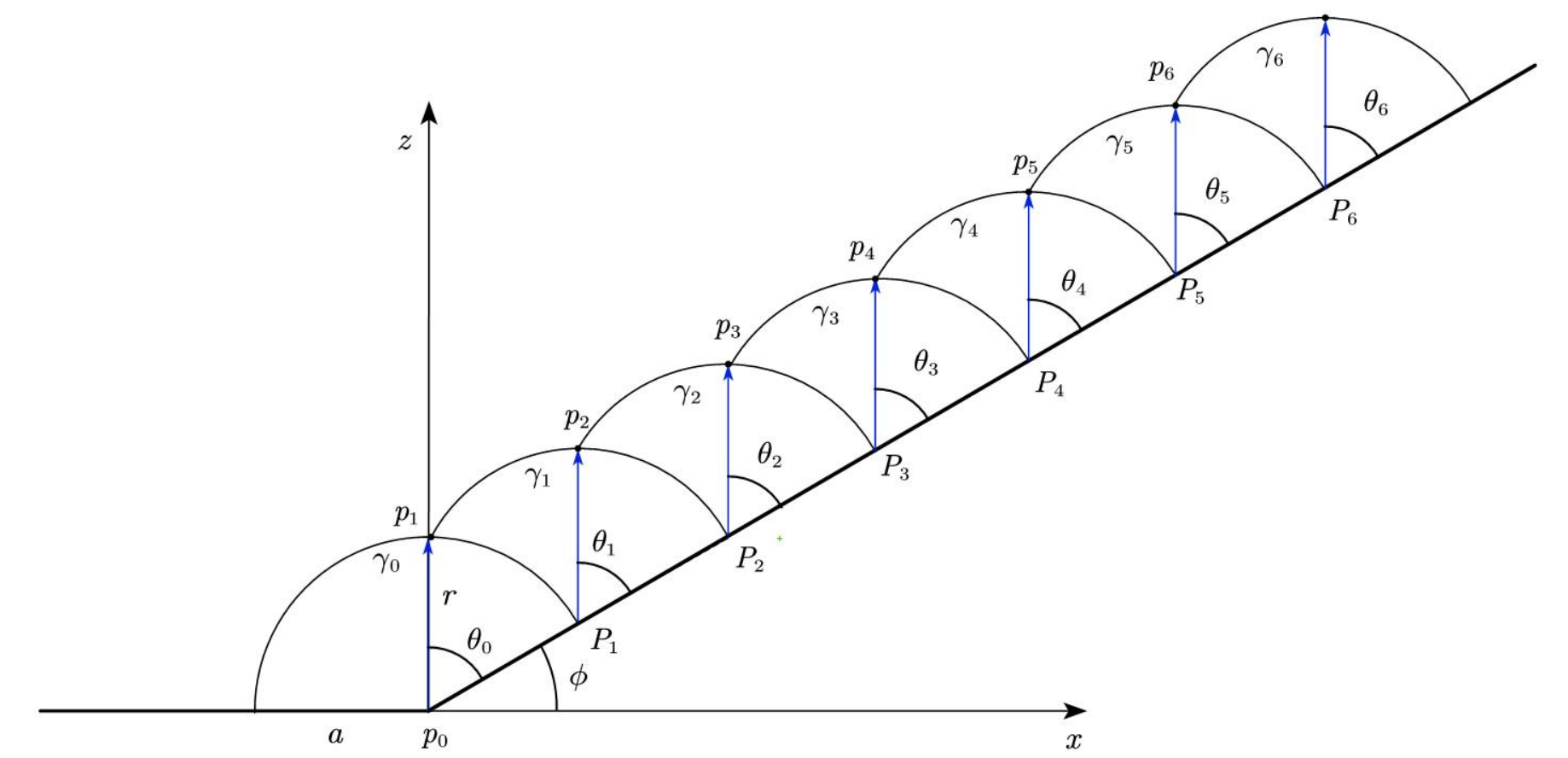}
	\caption{Entanglement wedge cross section in the growth scheme which contains seven layers of growing surfaces, $ \gamma_{i}$ denotes the extremal surface in the $i+1$-th layer, $ \theta_{i}$ is the location of wedge cross-section which are all equal to $ \frac{\pi}{3} $ and the boundary angle is $\phi =\frac{\pi}{6}$. The blue arrows denote the geodesics and indicate the direction of the entanglement of purification, and note that each geodesic circle has the same radius $r=a$. }
	\label{fig63}
\end{figure}

We can obtain a homogenous surface growth configuration in this case by choosing $\phi=\frac \pi 6$, then each geodesic circle has the same radius $r=a$. Then by iterating this process, one can obtain any number of layers of surface growth, for example, Fig.~\ref{fig63} illustrates the entanglement of purification in seven layers of homogenous surface growth. Besides, it is straightforward to see that when $\phi>\frac\pi 6$, the surface growth configurations cannot be homogenous and will decay after finite steps of surface growth. Since the entanglement of purification contains the minimal entanglement entropy in a given entanglement wedge, the intersection point $p_i$ naturally gives a fixed point and serves as the starting point of the next layer of growing surface, which gives more refined description of entanglement entropy in the entanglement wedge and indicates that the entanglement of purification plays a role in the dynamics of the surface growth for bulk reconstruction in the AdS/BCFT correspondence.

\section{Conclusion and Discussion}\label{conclusion}
In this paper, we extend the surface growth approach for bulk reconstruction into the AdS/BCFT correspondence. Our investigation mainly focus on the growth of strip-like bulk extremal surfaces in the presence of an additional spatial boundary, we show that the outer entanglement wedge can be constructed from the growth of two types of extremal surfaces, namely, the one which will attach the boundary while one will not, layer by layer, both in the homogenous and inhomogenous boundary subsystems.

In addition, we also show that the surface growth process in BCFT can correspond to the defect MERA tensor network by generalizing OSED tensor network into the situation with boundary or defect, which further demonstrated the connection between the emergence of bulk geometry and the entanglement renormalization of the tensor networks. Furthermore, since the surface growth scheme can divide a given entanglement wedge into arbitrary subregions, which naturally captures the fine structure of the entanglement entropy and reflects the correlations between these subregions, we study the entanglement of purification by investigating the entanglement wedge cross section in the surface growth scheme in AdS$_3$ spacetime with boundary and show that entanglement of purification can be determined in each step of the surface growth and can serve as a selection rule for the starting point of the next layer of extremal surface, which not only provides a more refined description of the entanglement entropy in the entanglement wedge but also might plays a role in the dynamics in the surface growth scheme for bulk reconstruction.

There are many interesting problems for further study, such as extending the surface growth scheme into boundary CFT with more complicated defect conditions or multiple boundaries, in which each defect described by the defect tensor network contains its corresponding causal wedge that will affect the surface growth process. It is also interesting to further study the surface growth scheme in higher dimensional BCFT case and the BCFT whose bulk duality contains a black hole.
\omits{
\begin{figure}
	\centering
	\centering
	\includegraphics[width=0.85\linewidth,height=10.5 cm,clip]{fig64.jpg}
	\caption{In the Type 2 scheme, the entanglement wedge cross-section contains five layers of growing minimal surfaces. The black dashed line is parallel to the $ x $ axis, connecting the center of the surface $ P_{i} $ and the last cross-section point $ p_{i-1}$, $ \gamma_{i}$ denot  RT surface in different steps, the boudary angle $\theta =\frac{\pi}{6}$, $ f_{i} $ is the intersection of corresponding surface with boundary,  blue  arrow denot the geodesic and the direction of entangement purification, the blach  line indicate  radius $ r_{i} $ of each different surface and the location of geodesic  sitting at each surface. }
	\label{fig64}
\end{figure}
}

\section*{Acknowledgement}
We would like to thank L.-Y. Hung for useful discussions. This project was supported by the National Natural Science Foundation of China (No.~11675272).

\end{document}